\documentstyle[amssymb,12pt]{article}
\def\beq{\begin{equation}}
\def\eeq{\end{equation}}
\def\bea{\begin{eqnarray}}
\def\eea{\end{eqnarray}}
\def\ba{\begin{array}}
\def\ea{\end{array}}

\def\ga{\gamma}

\def\om{\omega}

\def\tg{{\tilde\gamma}}
\def\rd{{\rm d}}
\def\ri{{\rm i}}

\def\ln{{\rm ln}}

\def\re{{\rm Re}}
\def\im{{\rm Im}}

\def\nn{\nonumber}

\begin{document}
\begin{center}
{\Large\bf Black hole entropy in Loop Quantum Gravity}

\vspace{0.5cm}
{\bf Krzysztof A. Meissner}

\vspace{0.5cm}
{\it
Institute of Theoretical Physics,
Warsaw University\\ Ho\.za 69,
00-681 Warsaw, Poland,\\}

\begin{abstract}
\noindent
We calculate the black hole entropy in Loop Quantum 
Gravity as a function of the horizon area and provide the exact formula 
for the leading and sub-leading terms. By comparison with the 
Bekenstein--Hawking formula we uniquely fix the value of the 
'quantum of area' in the theory.
\end{abstract}
\end{center}

The Bekenstein--Hawking formula gives the leading term in the entropy of 
the black hole in the form
\beq
S=\frac14\,\frac{A}{l_P^2}
\label{bekhaw}
\eeq
where $l_P$ is the Planck length and $A$ is the black hole horizon area. 
According to the common belief, the entropy is always connected with 
the logarithm of the number of microscopic states realizing a given 
macroscopic state. The fact that the entropy in (\ref{bekhaw}) is 
proportional to the area (and not as is usual to the volume) has led to 
the formulation of the so called 
holographic principle. It was however difficult to find 
the microscopic states that could account for such an entropy.

The problem was attacked in two different approaches -- in string theory 
\cite{tH}-\cite{DV} and in Loop Quantum Gravity 
\cite{LeeS}-\cite{ABK} 
(see also the review \cite{TT}). The latter approach is based on a 
quantum theory of geometry.
The basic geometric operators were introduced in \cite{RS,AL} and a
detailed description of the quantum horizon geometry was
introduced using the $U(1)$ Chern-Simons theory in \cite{ABK} where one 
can find the detailed discussion of the states on the horizon of 
the black hole (see also \cite{DL}). 
However, the procedure introduced in \cite{ABK} for state
counting contained a spurious constraint 
on admissible sequences and the number of the
relevant horizon states is underestimated. The correct method of
counting was proposed in \cite{DL} but that analysis provides only
lower and upper bounds on the number of states. The purpose 
of this paper is to rectify
this situation. We will provide the correct value of the
Barbero-Immirzi parameter that is needed to obtain agreement with
the Hawking-Bekenstein formula for large black holes. Furthermore,
since we are able to calculate the number of states to sufficient
accuracy, we also rigorously obtain the precise sub-leading
quantum correction to the Hawking-Bekenstein formula.
The physical aspects of this result will be discussed elsewhere 
\cite{ALM}.

As it was shown in \cite{DL} the proper counting of states is given 
by all sequences (of arbitrary length) of $m_i\in {\mathbb Z}/2,\ m_i\ne 
0$ such that
\beq
\sum_i\sqrt{|m_i|(|m_i|+1)}<a
\label{war1}
\eeq
where 
\beq
a=\frac{A}{8\pi\ga l_P^2}
\eeq
and $\ga$ is a parameter introduced in \cite{Imm}. 
Addditionally we have to impose the condition 
\beq
\sum_i m_i=0.
\label{war2}
\eeq
It is our task to calculate the number of sequences satisfying 
(\ref{war1}) and (\ref{war2}).

We start with the sequences without (\ref{war2}) imposed.
If we denote by $N(a)$ the number of sequences satisfying (\ref{war1}) 
then we can split the counting into the first number $m_1$ and the 
rest and write the recurrence relation
\bea
N(a)&=&\theta(a-\sqrt{3}/2)\left(2N(a-\sqrt{3}/2)+2 
N(a-\sqrt{2})+\ldots\right.\nn\\
&&
\left. +2 N(a-\sqrt{|m_i|(|m_i|+1})
+\ldots
+2\left[\sqrt{4a^2+1}-1\right]\right) 
\label{recrel}
\eea
where the symbol $[\ldots]$ denotes the integer part.
The first term on the RHS corresponds to sequences with at least two 
elements and $m_1=\pm 1/2$, the second to sequences with at least two
elements and $m_1=\pm 1$ and so on and the last term to 
sequences consisting of just $m_1$ and nothing else.

We now assume (we will later prove that it is indeed the case)
that the leading part of the solution of the recurrence relation 
(\ref{recrel}) is for large $a$ given by
\beq
N(a)=C e^{2\pi\ga_M a}
\label{Nassump}
\eeq
where $\ga_M$ is a constant to be determined. 

Plugging the form (\ref{Nassump}) into (\ref{recrel}) we get
\beq
1=\sum_{k=1}^{\infty} 2\,e^{-2\pi\ga_M\sqrt{k(k+2)/4}}
\label{sum12}
\eeq
where we neglected terms vanishing in the limit $a\to\infty$.
The solution to this equation can be found numerically and reads
\beq
\gamma_M=0.23753295796592\ldots
\label{defgam}
\eeq
We see that the coefficient of growth $\ga_M$ is between the lower and 
upper bounds given in \cite{DL} i.e. $\frac{\ln\,2}{\pi}<\ga_M< 
\frac{\ln\,3}{\pi}$. It seems that $\ga_M$ defined by (\ref{sum12}) 
cannot be calculated analytically or given as a solution of any 
``simpler'' equation. In the present paper we will also use $\tg$ where
\beq
\tg_M:=2\pi\ga=1.492463591462379\ldots
\label{deftg}
\eeq

Since the condition (\ref{war2}) (as we will later prove) gives 
only power-like corrections to $N(a)$ therefore for large $a$ we have 
from (\ref{Nassump}) 
\beq
S=\ln N(a)=\frac{\ga_M}{4\ga}\,\frac{A}{l_P^2} +O(\ln A).
\label{ent}
\eeq
The comparison with the Bekenstein--Hawking formula 
(\ref{bekhaw}) gives therefore the unambiguous result
\beq
\ga=\ga_M
\label{gagam}
\eeq
so this physical requirement uniquely fixes the value of $\ga$ and 
hence the 'quantum of area' in the theory.

One may note that the exponential ansatz
(\ref{Nassump}) works in many other cases for example if the number of
states for a given spin $j$ was $(2j+1)$ instead of $2$ one
would have in (\ref{sum12}) $(k+1)$ instead of $2$ in front of the
exponential and the appropriate $\ga_M$ would be equal to
$0.273985635\ldots$.

We now prove that the assumption (\ref{Nassump}) is actually 
correct and at the end we add the condition (\ref{war2}) -- the 
final result will not modify (\ref{gagam}) and will 
enable us to find the sub--leading (logarithmic) correction to the 
formula for entropy (\ref{ent}).

We introduce the Laplace transform of $N(a)$:
\beq
P(s):=\int_0^\infty \rd a\, N(a)\,e^{-s a}.
\label{defpa}
\eeq
The transform is well defined since we know that $N(a)$ is piecewise 
continuous and of exponential order i.e. it is bounded by $K e^{-\beta 
a}$ for some $K$ and $\beta$ (\cite{DL}). 
Integrating the relation (\ref{recrel}) over $a$  from $0$ to $\infty$ 
we get an exact result
\beq
P(s)
=\frac{2\sum_{k=1}^\infty e^{-s\sqrt{k(k+2)/4}}}
{s\left(1-2\sum_{k=1}^\infty e^{-s\sqrt{k(k+2)/4}}\right)}.
\label{ps}
\eeq
If all the poles of the  Laplace transform are distinct and of finite 
order then we can 
make analytical continuation to the complex half-plane $\re(s)>0$, 
excluding positions 
of the poles, i.e. beyond the region allowed by the definition 
(\ref{defpa}).   
We now use the well known fact that the growth of $N(a)$ is determined 
by the poles of $P(s)$ i.e.
\beq
N(a)=\sum_{s_i,\,\re(s_i)>0} {\rm res}_{s_i} e^{s_i a}+O(a^n)
\eeq
(we assumed here that all the poles are simple -- otherwise we would 
have to add some powers of $a$ in front of the exponentials). Therefore 
our problem boils down to determination of such complex $s_i$ that 
satisfy
\beq
G(s_i)-\frac12=0
\eeq
where we defined
\beq
G(s):=\sum_{k=1}^\infty e^{-s\sqrt{k(k+1)/4}}.
\eeq

It is important to analyze the distribution of zeroes of
$(G(s)-\frac12)$ in more detail. Since $G(s)$ is
complex analytic in the open domain $\re(s)>0$ all the zeroes are
distinct, of finite order and there is no
point of accumulation of zeroes inside this domain. Therefore we have to
analyze the domain's border i.e. the imaginary line $\re(s)=0$.
To analyze the limit when $\re(s)\searrow 0$ define a
new function $G_r(s)$ by subtracting two analytic (for $\re(s)>0$)
functions
\beq
G_r(s):=G(s)-\sum_{k=1}^{\infty}e^{-s(k+1)/2}-
\frac{s}{4}\sum_{k=1}^{\infty}\frac{e^{-s(k+1)/2}}{k}
\eeq
The function $G_r(s)$ is well defined in the whole domain $\re(s)\ge 0$,
continuous and doesn't have any singularities in this domain.
On the other hand two subtracted functions have a well defined limit
when $\re(s)\searrow 0$ so we can write
\beq
G(s)=G_r(s)+\frac{e^{-s/2}}{e^{s/2}-1}
-\frac{se^{-s/2}}{4}\,\ln\left(1-e^{-s/2}\right).
\eeq
We see that apart from singularities at $s=4\pi\ri n$, the function
$G(s)$ has a well defined limit when $\re(s)\searrow 0$ and in the
domain $\re(s)> 0$ it is complex analytic so
in the whole domain $\re(s)\ge 0$ there can be no points of accumulation
of zeroes -- therefore all zeroes of the function $G(s)-\frac12$  in 
the whole domain $\re(s)\ge 0$ must be distinct and of finite order.

The only real zero is given by (\ref{deftg}):
\beq
s=\tg_M.
\eeq
Close to this zero the function $P(s)$ behaves as
\beq
P(s)\sim \frac{C_M}{s-\tg_M}
\eeq
where
\beq
C_M=-\frac{1}{2\tg_M G'(\tg_M)}=0.509202564\ldots
\eeq
Therefore the leading behaviour for large $a$ is given by
\beq
N(a)=C_M e^{\tg_M a}
\label{Nlead}
\eeq 
as was to be shown.

The complex zeroes of the function $G(s)-\frac12$ for $\re(s)>1.$
and $|\im(s)|<100$ are listed below: 
\begin{center}
\begin{tabular}{c|c}  
$\re(s)$&$\im(s)$\\ 
\hline
1.49246359\ldots&0\\ 
1.22393017\ldots&$\pm$ 22.1530069\ldots\\
1.41016352\ldots&$\pm$ 35.9362749\ldots\\ 
1.30363023\ldots&$\pm$ 58.0472739\ldots\\ 
1.18587535\ldots&$\pm$ 71.8281531\ldots\\
1.17654302\ldots&$\pm$ 79.9673974\ldots\\ 
1.07191106\ldots&$\pm$ 87.4120423\ldots\\ 
1.21660746\ldots&$\pm$ 93.9713059\ldots\\ 
\hline
\end{tabular} 
\end{center} 
It is easy to prove that all the complex zeroes (i.e. poles of 
$P(s)$) with nonvanishing imaginary part must have smaller real part 
than $\tg_M$ so in comparison to the leading behaviour (\ref{Nlead}) 
they give exponentially small (and rapidly oscillating) contribution to 
$N(a)$. 

To illustrate the method in the cases where the number of sequences is
explicitly known let us give two examples. For sequences of 
half-integers with the
condition $\sum |m_i|<a$ the number of sequences is $3^{[2a]}-1$ and by
the method described above we get 
\beq
P(s)=\frac{2}{s\left(e^{s/2}-3\right)}
\eeq 
and indeed the real pole is equal to $2\,\ln\, 3$.  For sequences of
half-integers with the condition $\sum (|m_i|+1/2)<a$ the number of
sequences is $\frac13(2^{[2a]+1}\pm 1)-1$ (where $+$ is for even and $-$
for odd $[2a]$) and 
\beq
P(s)=\frac{2}{s\left(e^{s}-e^{s/2}-2\right)}  
\eeq 
and indeed the positive real pole is equal to $2\,\ln\, 2$. 
The actual numbers of sequences in both cases indeed show
that the leading behaviour is given precisely by the largest positive
real pole of $P(s)$.

It can be useful to discuss how different ways of defining the sequences 
are reflected in the properties of $P(s)$. We have essentially three 
different ``statistics''. 
\begin{itemize}
\item
Boltzmann statistics (used in this paper): all sequences satisfying the 
condition are allowed and are treated as different. Therefore $P(s)$ is 
given by eq. (\ref{ps}): 
\beq
P(s)
=\frac{2\sum_{k=1}^\infty e^{-s\sqrt{k(k+2)/4}}}
{s\left(1-2\sum_{k=1}^\infty e^{-s\sqrt{k(k+2)/4}}\right)}
\eeq
which has the largest positive real pole equal to $\ga_M$.
\item
the intermediate statistics (between Boltzmann 
and Bose): it requires that $|m_{i+1}|\ge |m_i|$ but treats for example 
$(+\frac12,-\frac12,+\frac12,\ldots)$ as different from 
$(-\frac12,+\frac12,+\frac12,\ldots)$ -- then we have 
\beq
P(s)
=\prod_{k=1}^\infty \frac{1}
{\left(1-2e^{-s\sqrt{k(k+2)/4}}\right)}
\eeq
which has the largest positive real pole given by $k=1$: 
$s=\frac{2\ln\,2}{\sqrt{3}}$.
\item
Bose statistics: it requires not only that 
$|m_{i+1}|\ge |m_i|$ but treats for example
$(+\frac12,-\frac12,+\frac12,\ldots)$ as the same as
$(-\frac12,+\frac12,+\frac12,\ldots)$ i.e. the allowed sequences have 
the form $(\frac12,\ldots,-\frac12,\ldots,+1,\ldots,-1,\ldots)$. Then we 
have
\beq
P(s)
=\prod_{k=1}^\infty \frac{1}
{\left(1-e^{-s\sqrt{k(k+2)/4}}\right)^2}
\eeq
and this $P(s)$ does not have a pole with $\re(s)>0$ at all so $N(a)$
must have milder growth than $e^{c'a}$. However when $s\to 0$ the 
function $P(s)$ behaves as $P(s)\sim e^{\frac{2\zeta(2)}{s}}$ where 
$\zeta(s)$ is the Riemann's zeta function. Hence we have 
accumulation of zeroes at $s=0$ and the inverse Laplace transform
gives the exponential growth with $\sqrt{a}$:
\beq
N(a)\sim a^c\,e^{\sqrt{8\zeta(2)a}}
\eeq
(this type of statistics and therefore similar formula gives for 
example the degeneracy of states in string theory). 
\end{itemize}

At the end we impose the condition (\ref{war2}) i.e. $\sum m_i =0$. 
We introduce $N(a,p)$ as the number of sequences satisfying  
\beq
\sum_i\sqrt{|m_i|(|m_i+1)}<a
\eeq
and 
\beq
\sum_i m_i=p
\eeq
where $p\in{\mathbb Z}/2$ and $a\ge\sqrt{|p|(|p|+1)}$.
We can write the recurrence relation as
\bea
N(a,p)&=&\theta(a-\sqrt{3}/2)\theta(a-\sqrt{|p|(|p|+1)})\times\nn\\
&&\left(N(a-\sqrt{3}/2,p-1/2)+(N(a-\sqrt{3}/2,p+1/2)+\right.\nn\\
&&
\left. +N(a-\sqrt{2},p-1)+N(a-\sqrt{2},p+1)+\ldots +1\right).
\label{recrelb}
\eea
where the first term inside the brackets on the RHS corresponds to 
sequences with at least two
elements and $m_1=+\frac12$, the second to sequences with at least two
elements and $m_1=-\frac12$ and so on and the last term to
sequences consisting of just $m_1$ (which must be equal to $p$).

Next we define $P(s,\om)$ as the Fourier transform with 
respect to $p$ and Laplace transform with respect to $a$ of $N(a,p)$:
\beq
P(s,\om):=\sum_{p\in{\mathbb Z}/2} \int_0^\infty\rd a
\,N(a,p)\,e^{\ri\om p}\, e^{-s a} .
\eeq
Then summing and integrating eq. (\ref{recrelb}) we get
\beq
P(s,\om)
=\frac{2}{s}\sum_{k=1}^\infty e^{-s\sqrt{k(k+2)/4}}
\left(1-2\sum_{k=1}^\infty
e^{-s\sqrt{k(k+2)/4}}\cos(k\om/2)\right)^{-1}
\label{palom}
\eeq
where we neglected some contributions from the endpoints 
($p\sim a$) what will be justified {\it a posteriori}. Therefore we 
have to find the zeroes of the function inside the bracket on the RHS of 
(\ref{palom}). With $\om=0$ we know that the only 
real zero (i.e. the one that gives the leading 
behaviour) is equal to $s=\tg_M$. 
Expanding around $s=\tg_M$ we get
\beq
1-2\sum_{k=1}^\infty e^{-\tg_M\sqrt{k(k+2)/4}}
\left(1-(s-\tg_M)\sqrt{k(k+2)/4}-k^2\om^2/8+O(\om^4)\right)=0.
\eeq
Hence the pole of $P(s,\om)$ is given by
\beq
s=\tg_M-\beta_M\om^2 +O(\om^4)
\label{alzerom}
\eeq
where
\beq
\beta_M=-\frac{\sum_{k=1}^\infty k^2 
e^{-\tg_M\sqrt{k(k+2)/4}}}{8G'(\tg_M)}
=0.475841255\ldots
\eeq
Therefore the leading behaviour of $N(a,p)$ is given by
\beq
N(a,p)=\frac{C_M}{2\pi}\int\rd\om\, e^{-i\om p}\,e^{\tg_M a-\beta_M\om^2 
a}=\frac{C_M}{\sqrt{4\pi\beta_M a}}\, e^{2\pi\ga_M a}\,e^{-p^2/(4\beta_M 
a)}.
\label{Nap}
\eeq
As we see the endpoints $p\sim a$ are exponentially small justifying 
(\ref{palom}). Also we see {\it a posteriori} that the terms 
$O(\om^4)$ from (\ref{alzerom}) would give in (\ref{Nap})  
corrections of relative size $1/a$ so for large $a$ we can neglect them.

Note that the gaussian distribution with respect 
to $p$ in (\ref{Nap}) with the mean square deviation of the order of 
$\sqrt{a}$ can be expected by the analogy with the random walk where 
the average distance is zero and the  mean square deviation is of the 
order of square root of the number of steps.   

Imposing the condition $p=0$ we get the final result
\beq
N(a,0)=\frac{C_M}{\sqrt{4\pi\beta_M a}}\, e^{2\pi\ga_M a}.
\label{final}
\eeq
Therefore the entropy is given by
\beq
S=\ln N(a,0)=\frac{\ga_M}{4\ga}\,\frac{A}{l_P^2}-\frac{\ln 
(A/l_P^2)}{2} 
+O(1)
\label{ent2}
\eeq
so for large $A/l_P^2$ we confirm (\ref{ent}) as the leading behaviour
and we can unambiguously both get the physical value of 
$\ga$ equal to $\ga_M$ (given in (\ref{defgam})) and 
resolve the controversy (summarized in \cite{GM}) about the 
coefficient of the logarithmic term fixing it to $-\frac12$.

\medskip

\noindent {\bf Acknowledgement}
The author would like to acknowledge valuable discussions with 
Abhay Ashtekar and Jerzy Lewandowski and thank Hermann Nicolai for 
the invitation to the Albert Einstein Institute in Golm where part of
this work was done. The work partially supported by the Polish KBN 
grant 2P03B 001 25 and the European Programme HPRN--CT--2000--00152.

\vspace{0.5cm}
{\it E-mail address:} {\tt Krzysztof.Meissner@fuw.edu.pl}

\end{document}